\documentclass{article}
\usepackage{amssymb}
\usepackage{amsmath}

\input{tcilatex}

\begin{document}

\section{introduction}

Model considerations are always very important for the concept clarification
in theoretical and experimental physics. In particular, an exact solvable
model may provide much more detailed informations which are useful for a
complete understanding of the nature of the phenomena. Thus, in the study of
magnetic phase transitions of inhomogeneous systems, different ways of
modeling the systems can be found in the literature. But to implement the
inhomogeneity of real systems into a model usually make the model\ itself
hardly to have exact solution. However, inhomogeneous systems are frequently
encountered in nature, and they also play an important role in material
science. For example, to develope new materials, the inhomogeneous systems
formed by doping with other materials in a bulk system are often grown
artificially in laboratory. Thence, it is important to construct some exact
solvable models so that we can deepen our insights to the nature of such
systems. The hierarchical spin models may be viewed as one of the efforts in
this direction.

About one and half decades ago, spin models defined on hierarchical lattices
received much attention in the literature$\left[ 1-9\right] $. Hierarchical
lattices are defined as the infinite limit of iterative steps of replacing
each fundamental bond or cell with a given type of decoration unit, and
thence they are self-similar fractal lattices. Since hierarchical lattices
are highly inhomogeneous in the local connectivity and lack of translational
invariance, spin models defined on hierarchical lattices may provide a very
good frame in understanding the magnetic properties of inhomogeneous
systems. Moreover, we can apply the Migdal-Kadanoff renormalization$\left[
10,11\right] $ scheme to obtain exact solutions of the partition functions
of hierarchical spin models. Among a variety of hierarchical spin models,
the zero-field $q$-state Potts model defined on diamond-hierarchical lattice
has been widely studied. The results revealed from the exact solutions show
that the corresponding systems have some nonclassical thermodynamic
features. They include that the second-order phase transition occurs at
finite temperature with the cusp-behavior appearing in the specific heat at
the critical point$\left[ 8,9,12\right] $, the distribution of the partition
function zeros owns a multifractal structure on the complex temperature
plane, and the free energy near the critical point has a spatially modulated
structure$\left[ 6,7\right] $.

We notice that owing to the infinite iterative steps of decorations in
lattice construction, there exist a well-defined thermodynamic limit for a
hierarchical spin model. Thence, the degree of the inhomogeneity is
completely specified for a hierarchical spin model. If we had performed any
finite iterative steps of decorations in lattice constrution, it would lead
to the lack of thermodynamic limit for the model. To improve this while
remain the model to be exactly solvable, we may study Ising systems defined
on decorated lattices.

A decorated lattice is constructed by implementing bond- or cell-
decorations to a regular lattice in an iterative way. The geometric
complexity of a decorated lattice then is characterized by the decoration
level $n$. For a decorated lattice of the decoration level $n$, a primary
bond or cell defined on the original regular lattice is replaced by an $n$%
-bond or cell. An $n$-bond or cell owns a complicated inherent structure
formed by $n$ iterative steps of replacing a bond or cell with some basic
unit of decoration. In the limit of infinite $n$, an $n$-bond or cell
essentially possesses a fractal lattice as the inherent structure. For this
type of lattices, the degree of inhomogeneity in the local connectivity is
proportional to the decoration level $n$ and there exists a well-defined
thermodynamic limit for any value of $n$.

Triangular type decorated Ising models are examples. The analyses on these
models have been carried out by Plechko$\left[ 13\right] $ and Plechko and
Sobolev$\left[ 14\right] $, and there exists some non-classical results. In
particular, the results of $\left[ 14\right] $ about the specific heat of
the Ising model on cell-decorated triangular lattices have revealed some
unusual features of ferromagnetic phase transitions. In the curve of the
specific heat versus the temperature, as $n$ increases, the critical region
is narrowing and a round peak appears above the critical point. While the
critical point marks the set in for the long-range ordering among different $%
n$-cells, the appearence of the round peak corresponds to the occurrence of
the short-range ordering inside an $n$-cell. As $n$ approaches to the limit
of infinity, an $n$-cell has the Sierpi{\'{n}}ski gasket as its inherent
structure. Then, in approaching the limit, $n$-cells are factorized and
become independent in the Ising system. Moreover, the critical point
approaches the zero temperature, and the round peak temperature remains
finite.

Bond-decorated and cell-decorated lattices are different ways of modelling
inhomogeneous systems. As the decoration level $n$ increases, the order of
ramification remains a constant for a cell-decorated triangular type lattice
but it increases unboundedly for a bond-decorated square lattice.
Furthermore, it was pointed out by Gefen et. al. that there is no phase
transition at finite temperature for a fractal lattice with the finite order
of ramification$\left[ 15\right] $. Hence, the non-universal features of
phase transitions in cell-decorated Ising models may be quite different from
that of bond-decorations. To obtain a more complete picture about the
possible features of inhomogeneous systems, we study the diamond-type
bond-decorated Ising model. Note that the diamond-type bond-decoration was
also adopted in a widely studied model, the diamond-hierarchical $q$-state
Potts model$\left[ 1-7\right] $. In this work, we are interested in the
following questions: $\left( i\right) $ How does the degree of the
inhomogeneity in a system, characterized by the decoration level $n$, affect
the nature of the phase transition? $\left( ii\right) $ How do the
characteristic features of ferromagnetic phase transitions change with the
lattice structure varying from cell-decorated triangular type lattices to
bond-decorated square lattices? $\left( iii\right) $ How do the
non-universal features obtained from the diamond-hierarchical Ising model$%
\left[ 1-9\right] $ appear? In particular, can we understand better about
the cusp-behavior appearing in the specific heat at the critical point
instead of the well-known logarithmic divergence?

The analytic expression of the partition function of square Ising model with
diamond-type bond-decorations can be obtained by referring to the exact
result of two-dimensional Ising model. The exact solution of the Ising model
on a square lattice was first given by Onsager$\left[ 16\right] $. Since
then, other alternatives in obtaining the exact solution have been
developed. Besides of the combinatorial method, which was first developed by
Kac and Ward$\left[ 17\right] $ and then rigorously reformulated by Green
and Hurst$\left[ 18\right] $, more recently Plechko used a nonstandard
method of Grassman multipliers to obtain the analytic expressions of the
partition functions for a variety of cases$\left[ 13,14,19-21\right] $.
Based on the frame provided by these results, in Section 2 we use the
bond-renormalization scheme to construct the analytic solution of the free
energy. In Section 3, we determine the critical point of an $n$-lattice and
study how the critical temperature vary with $n$. The condition of the
critical point is given by the occurrence of the non-analyticity in the free
energy. In Section 4, we use the solution of the free energy to calculate
the internal energy and specific heat in an analytic way. Since the free
energy can be expressed as the sum of those contributed by the local
interactions among the sub-bonds of an $n$-bond and the long-range
interactions among different $n$-bonds, we also use the same separation in
the internal energy and specific heat. Thence, in general, we can obtain
some insights about the nature of the phase transition of an inhomogeneous
system by comparing the contributions from the local and the long-range
interactions. Our results indicate that the cross over from a
finite-decorated system to an infinite-decorated system may not be a smooth
continuation, we then calculate the critical values of the internal energy
and specific heat of an infinite-decorated system in Section 5. Finally,
Section 6 is preserved for the summary and discussion.

\section{free energy}

\label{s2}We study the Ising model defined on an arbitrary $n$-lattice. A
two-dimensional square lattice is used as backbones for the construction of
an $n$-lattice. The simple square lattice is referred as $0$-lattice, and
its connecting bonds between lattice sites are named as $0$-bonds. For a $0$%
-lattice, the total site-number is denoted as $n_{s}$ and the total
bond-number as $n_{b}$ with $n_{b}=2n_{s}$. Starting from a $0$-bond, we
obtain the corresponding $n$-bond as the result of replacing a bond with a
diamond iteratively $n$ times. This $n$-bond then is specified by the two
sites connected by the $0$-bond. For an $n$-bond, we have the total
site-number $s^{\left( n\right) }=2(4^{n}+2)/3$ and bond number $b^{\left(
n\right) }=4^{n}$. The construction of an $n$-bond is schematized in Fig. 1.
From the primary structure of a $0$-lattice, we replace each $0$-bond with
an $n$-bond to form an $n$-lattice. For an $n$-lattice, the average site-
and bond-numbers per unit square are $N_{s}^{\left( n\right) }=2s^{\left(
n\right) }-3$ and $N_{b}^{\left( n\right) }=2b^{\left( n\right) }$
respectively. Examples of $n$-lattices are shown in Fig. 2.

The general form of the partition function for the Ising model defined on an 
$n$-lattice reads 
\begin{equation}
Z^{\left( n\right) }=\sum\limits_{\left\{ \sigma \right\} }\left[
\prod_{\left\langle i,j\right\rangle }\exp \left( \eta \sigma _{i}\sigma
_{j}\right) \right] ,
\end{equation}
where the sum is over all bond-connected pairs $\left\langle
i,j\right\rangle $ of $\ $the $n$-lattice, and the Ising spin takes two
possible values $\sigma _{i}=\pm 1$. Here we consider uniform ferromagnetic
couplings characterized by the coupling strength $J$, and the dimensionless
coupling parameter $\eta $ is defined as $\eta =J/k_{B}T$.

To implement the bond-renormalization scheme in the evaluation of the
partition function, we rewrite Eq. (1) as 
\begin{equation}
Z^{\left( n\right) }=2^{n_{b}\left( s^{\left( n\right) }-2\right)
}\sum\limits_{\left\{ \sigma \right\} }\left[ \prod_{\left\langle \mu
,\upsilon \right\rangle }B_{\left\langle \mu ,\upsilon \right\rangle
}^{\left( n\right) }\right] ,
\end{equation}
where the Boltzmann $n$-factor, $B_{\left\langle \mu ,\upsilon \right\rangle
}^{\left( n\right) }$, is the contribution to the partition function from
the Boltzmann factors associated with the $n$-bond specified by the two
sites, $\mu $ and $\upsilon $, the sum is over the spins defined only on the
lattice sites of the corresponding $0$-lattice, and the front factor is
added to compensate the normalization factor added in $B_{\left\langle \mu
,\upsilon \right\rangle }^{\left( n\right) }$. Note that the number of the
decorated sites of an $n$-bond is $s^{\left( n\right) }-2$, and the Ising
spins on the decorated sites are referred as the inner-spins which couple
only to the spins on the same $n$-bond. Thence, the $B_{\left\langle \mu
,\upsilon \right\rangle }^{\left( n\right) }$ of Eq. (2) is defined as the
result of taking the sum over the inner-spins for the product of all
Boltzmann factors associated with the $n$-bond, 
\begin{equation}
B_{\left\langle \mu ,\upsilon \right\rangle }^{\left( n\right) }=\left( 
\frac{1}{2}\right) ^{\left( s^{\left( n\right) }-2\right) }\sum_{\left\{
\sigma _{k},k=1,...,s^{\left( n\right) }-2\right\} }\exp \left[ \eta \left(
\sigma _{\mu }\sigma _{1}+\sigma _{1}\sigma _{2}+...+\sigma _{s^{\left(
n\right) }-2}\sigma _{\nu }\right) \right] ,
\end{equation}
where the sum is over the $\left( s^{\left( n\right) }-2\right) $
inner-spins, and the front factor is added for the normalization of the sum.

To obtain the explicit form of $B_{\left\langle \mu ,\upsilon \right\rangle
}^{\left( n\right) }$ for an arbitrary $n$, first we express the Boltzmann $%
0 $-factor, which takes the form of $\exp \left( \eta \sigma _{\mu }\sigma
_{\upsilon }\right) $, as 
\begin{equation}
B_{\left\langle \mu ,\upsilon \right\rangle }^{\left( 0\right) }=\left(
\cosh \eta \right) +\sigma _{\mu }\sigma _{\upsilon }\left( \sinh \eta
\right) .
\end{equation}
According to the definition of Eq. (3), the Boltzmann $1$-factor is given as 
\begin{equation}
B_{\left\langle \mu ,\upsilon \right\rangle }^{\left( 1\right) }=\left( 
\frac{1}{2}\right) ^{2}\sum_{\sigma _{1},\sigma _{2}}\exp \left[ \eta \left(
\sigma _{\mu }\sigma _{1}+\sigma _{\mu }\sigma _{2}+\sigma _{1}\sigma _{\nu
}+\sigma _{2}\sigma _{\nu }\right) \right] .
\end{equation}
Using Eq. (4) for $\exp \left( \eta \sigma _{\mu }\sigma _{\upsilon }\right) 
$, after some algebraic manipulations we obtain the result of Eq. (5) as 
\begin{equation}
B_{\left\langle \mu ,\upsilon \right\rangle }^{\left( 1\right) }=\left[ \exp
(\eta ^{\left( 1\right) })\right] \left[ \cosh \left( \eta ^{\left( 1\right)
}\right) +\sigma _{\mu }\sigma _{\upsilon }\sinh \left( \eta ^{\left(
1\right) }\right) \right] ,
\end{equation}
with 
\begin{equation}
\exp \left( \eta ^{\left( 1\right) }\right) =\cosh \left( 2\eta \right) .
\end{equation}
By continuing such construction, we can express Eq. (3) as 
\begin{equation}
B_{\left\langle \mu ,\upsilon \right\rangle }^{\left( n\right) }=R^{\left(
n\right) }\left( \eta \right) \left[ \cosh \left( \eta ^{\left( n\right)
}\right) +\sigma _{\mu }\sigma _{\upsilon }\sinh \left( \eta ^{\left(
n\right) }\right) \right]
\end{equation}
for $n\geq 1$, where the function $R^{\left( n\right) }\left( \eta \right) $
is given by 
\begin{equation}
R^{\left( n\right) }\left( \eta \right) =\prod_{k=1}^{n}\left[ \exp \left(
\eta ^{\left( k\right) }\right) \right] ^{4^{n-k}},
\end{equation}
and $\eta ^{\left( k\right) }$ is the image of $\eta ^{\left( k-1\right) }$
under the renormalization map defined as 
\begin{equation}
\exp \left( \eta ^{\left( k\right) }\right) =\cosh \left( 2\eta ^{\left(
k-1\right) }\right)
\end{equation}
with the initial condition, $\eta ^{\left( 0\right) }=\eta $ for $1\leq
k\leq n$.

For the Ising system defined on a single $n$-bond, the partition function of
Eq. (2) has $n_{b}=1$ and it reduces to 
\begin{equation}
Z_{b}^{\left( n\right) }=2^{s^{\left( n\right) }}\left( \frac{1}{2}\right)
^{2}\sum_{\sigma _{\mu },\sigma _{\upsilon }}B_{\left\langle \mu ,\upsilon
\right\rangle }^{\left( n\right) }.
\end{equation}
After the substituition of Eq. (8) for $B_{\left\langle \mu ,\upsilon
\right\rangle }^{\left( n\right) }$, this yields 
\begin{equation}
Z_{b}^{\left( n\right) }=2^{s^{\left( n\right) }}R^{\left( n\right) }\left(
\eta \right) \cosh \left( \eta ^{\left( n\right) }\right) .
\end{equation}
Then, the corresponding free energy per bond per $k_{B}T$, is given as 
\begin{equation}
f_{b}^{\left( n\right) }=-\frac{1}{b^{\left( n\right) }}\ln Z_{b}^{\left(
n\right) },
\end{equation}
where $b^{\left( n\right) }$ is the total bond number contained in an $n$%
-bond. This, up to a constant, yields 
\begin{equation}
f_{b}^{\left( n\right) }=-\frac{1}{4^{n}}\ln \left[ \cosh \left( \eta
^{\left( n\right) }\right) \right] -\sum_{k=1}^{n}\frac{\eta ^{\left(
k\right) }}{4^{k}}.
\end{equation}
Note that $f_{b}^{\left( n\right) }$ in the limit of infinite $n$ is exactly
the free energy density of the diamond-hierarchical Ising model.

Using the result of Eq. (12) for $Z_{b}^{\left( n\right) }$, we can express
Eq. (2) in terms of $Z_{b}^{\left( n\right) }$ as 
\begin{equation}
Z^{\left( n\right) }=2^{n_{s}}\left( Z_{b}^{\left( n\right) }\right)
^{n_{b}}Q^{\left( n\right) },
\end{equation}
where $Q^{\left( n\right) }$, referred as the reduced partition function,
takes the form of 
\begin{equation}
Q^{\left( n\right) }=\ (\frac{1}{2})^{n_{s}}\sum\limits_{\left\{ \sigma
\right\} }\left\{ \prod\limits_{\left\langle \mu ,\upsilon \right\rangle }%
\left[ 1+\sigma _{\mu }\sigma _{\upsilon }\tanh \eta ^{\left( n\right) }%
\right] \right\} .
\end{equation}

The reduced partition function of an $n$-lattice given by Eq. (16) has
exactly the same form as that of simple square lattice when the variable $%
\eta ^{\left( n\right) }$ is changed to $\eta $. Thence, the corresponding
free energy can be easily written down according to the formal expression of
exact solution of simple square lattice provided by Refs. $\left[
13,16,19,21,22\right] $. Up to a constant, we can express the free energy
per bond per $k_{B}T$ as 
\begin{equation}
f^{\left( n\right) }=f_{b}^{\left( n\right) }+f_{bb}^{\left( n\right) },
\end{equation}
where $f_{b}^{\left( n\right) }$ is given by Eq. (14), and $f_{bb}^{\left(
n\right) }$ is the contribution from $Q^{\left( n\right) }$ and it is given
as 
\begin{equation}
f_{bb}^{\left( n\right) }=\frac{-1}{2N_{b}^{\left( n\right) }}\int_{0}^{2\pi
}\frac{d\phi }{2\pi }\int_{0}^{2\pi }\frac{d\theta }{2\pi }\ln \left[
A_{0}^{(n)}-A_{1}^{(n)}\left( \cos \theta +\cos \phi \right) \right] ,
\end{equation}
with $A_{0}^{(n)}$ and $A_{1}^{(n)}$ defined as 
\begin{equation}
A_{0}^{(n)}=\frac{\cosh ^{2}\left( 2\eta ^{\left( n\right) }\right) }{\cosh
^{4}\left( \eta ^{\left( n\right) }\right) },
\end{equation}
\begin{equation}
A_{1}^{(n)}=\frac{2\sinh \left( \eta ^{\left( n\right) }\right) }{\cosh
^{3}\left( \eta ^{\left( n\right) }\right) }.
\end{equation}
Since $f_{b}^{\left( n\right) }$ is the free energy density of independent $%
n $-bonds, $f_{bb}^{\left( n\right) }$ can be identified as the contribution
from the interactions among different $n$-bonds. Thus, we refer $%
f_{b}^{\left( n\right) }$ as the contribution from the \textit{local
interactions} and $f_{bb}^{\left( n\right) }$ as that from the \textit{%
long-range interactions}.

To simplify the expression of $f_{bb}^{\left( n\right) }$, we change the
variables, $\omega _{1}=\left( \theta +\phi \right) /2$ and $\omega
_{2}=\left( \theta -\phi \right) /2$, to rewrite Eq. (18) as 
\begin{equation}
f_{bb}^{\left( n\right) }=-\frac{1}{2N_{b}^{\left( n\right) }}\ln
A_{0}^{(n)}-\frac{1}{N_{b}^{\left( n\right) }}\int_{0}^{\pi /2}\frac{d\omega
_{2}}{\pi }\int_{0}^{\pi }\frac{d\omega _{1}}{\pi }\ln \left( 1-\kappa
^{\left( n\right) }\cos \omega _{1}\cos \omega _{2}\right) ,
\end{equation}
with 
\begin{equation}
\kappa ^{\left( n\right) }=\frac{2A_{1}^{(n)}}{A_{0}^{(n)}}.
\end{equation}
Then, the integral over $\omega _{1}$ can be easily performed and the result
is 
\begin{equation}
f_{bb}^{\left( n\right) }=-\frac{1}{2N_{b}^{\left( n\right) }}\ln
A_{0}^{(n)}-\frac{1}{N_{b}^{\left( n\right) }}\int_{0}^{\pi /2}\frac{d\omega 
}{\pi }\ln \left[ \frac{1}{2}\left( 1+\sqrt{1-\kappa ^{\left( n\right)
2}\sin ^{2}\omega }\right) \right] .
\end{equation}

\section{critical point}

The free energy density, $f_{bb}^{\left( n\right) }$, is not completely
analytic for the physical region of temperature. The non-analyticity arises
from the second derivative of $f_{bb}^{\left( n\right) }$ with respect to
temperature at the point $\kappa ^{\left( n\right) }=1$ for the physical
region $0\leq \kappa ^{\left( n\right) }\leq 1$. This implies that for the
Ising model on an $n$-lattice, the long-range interactions among different $%
n $-bonds are responsible for the occurrence of the Ising phase transition
at the finite temperature, and the critical point is determined by the
condition $\kappa ^{\left( n\right) }=1$. Thus, we can obtain the bulk
critical temperature of the ferromagnetic phase transition from the solution
of the condition, 
\begin{equation}
A_{0}^{(n)}-2A_{1}^{(n)}\overset{c}{=}0.
\end{equation}
Here, for convenience, we use the notation, $\overset{c}{=}$, to denote the
equivalence established only at the critical temperature. With the
definitions of $A_{0}^{(n)}$ and $A_{1}^{(n)}$ given by Eqs. (19) and (20),
we can rewrite the critical condition as the familiar form, $\sinh 2\eta
^{\left( n\right) }\overset{c}{=}1$, or as 
\begin{equation}
\exp \left( \eta ^{\left( n\right) }\right) \overset{c}{=}h^{(0)}
\end{equation}
with 
\begin{equation}
h^{(0)}=\sqrt{1+\sqrt{2}}
\end{equation}
for an $n$-lattice.

As the consequence of the map of Eq. (10), Eq. (25) yields the critical
value of the $\exp \left( \eta ^{\left( n-1\right) }\right) $ variable as 
\begin{equation}
\exp \left( \eta ^{\left( n-1\right) }\right) \ \overset{c}{=}h^{(1)},
\end{equation}
for $n\geq 1$ with 
\begin{equation}
h^{(1)}\ =\ \left( h^{\left( 0\right) }+\sqrt{\left( h^{(0)}\right) ^{2}-1}%
\right) ^{1/2}.
\end{equation}
By applying the recursion relation of Eq. (10) continuously, we can
establish the result, 
\begin{equation}
\exp \left( \eta ^{\left( n-k\right) }\right) \ \overset{c}{=}h^{(k)},
\end{equation}
for $1\leq \ k\leq n$, where $h^{\left( k\right) }$ is determined by the
recursion relation 
\begin{equation}
h^{(k)}\ =\ \left( h^{\left( k-1\right) }+\sqrt{\left( h^{(k-1)}\right)
^{2}-1}\right) ^{1/2}.
\end{equation}

By taking $k=n$ in Eq. (29) we obtain the reduced critical temperature of
the Ising transition as 
\begin{equation}
\left( \frac{k_{B}T_{c}}{J}\right) _{n}=\left[ \ln \left( h^{\left( n\right)
}\right) \right] ^{-1}
\end{equation}%
for an arbitrary $n$-lattice. For $n=0$, this gives the well-known result, $%
k_{B}T_{c}/J=2.269185$. As $n$ approaches $\infty $, the $h^{\left( n\right)
}$ value can be determined by the fixed points of the map of Eq. (30). In
fact, the $h^{\left( \infty \right) }$ value is the repellor of the map of
Eq. (30), and this repellor locates at $h^{\left( \infty \right) }=1.839287$%
. The corresponding critical temperature is $\left( k_{B}T_{c}/J\right)
_{n\rightarrow \infty }=1.641018$. However, the nature of the phase
transition at the repellor remains to be answered, and this will be
discussed in Section 5 where the case of infinite decoration-level is
discussed.

The numerical values of $\left( k_{B}T_{c}/J\right) _{n}$ for some different 
$n$ values are given in Table 1, and the result as a function of $n$ is
shown in Fig. 3 with the solid curve given by the fitting function $\left(
k_{B}T_{c}/J\right) _{n}=1.6410+\left( 0.6281\right) \exp \left[ -\left(
0.5857\right) n\right] $. This result indicates that the critical
temperature decreases exponentially toward the limiting value $1.6410$ as $n$
increases. On the other hand, it has been shown$\left[ 14\right] $ that for
the cell-decorated triangular Ising model the critical temperature as a
function of $n$ is given by $\left( k_{B}T_{c}/J\right) _{n}=4/\ln \left\{
4n-4\ln \left[ \left( n+1\right) /2\right] +1\right\} $ for $n\geq 1$, and
this leads to $\left( k_{B}T_{c}/J\right) _{n}=4/\ln \left( 4n\right) $ as $%
n\rightarrow \infty $. Thus, the decreasing rate of the critical temperature
toward the limit of infinite decoration level is more faster for the
bond-decorated Ising model in comparing with the cell-decorated Ising model.

\section{specific heat}

\label{s3} In this section, we follow the standard procedure, for example as
given by Ref. $\left[ 22\right] $, to calculate the specific heat for an $n$%
-lattice. The analytic result is expressed in terms of the complete elliptic
integrals of the first and second kinds. The contributions to the specific
heat contain those from the local and the long-range interactions, and the
comparison between these two is given quantitatively. The critical behaviors
of the specific heats of different $n$-lattices are also discussed. In
particular, we compare the results with the cell-decorated case.

Using the free energy density $f^{\left( n\right) }$ of an $n$-lattice given
by Eq. (17), we define the dimensionless internal energy per bond as 
\begin{equation}
\epsilon ^{\left( n\right) }=\frac{\partial f^{\left( n\right) }}{\partial
\eta }.
\end{equation}
Due to the separable contributions to the free energy density, we can write
the internal energy density as the sum of two parts, 
\begin{equation}
\epsilon ^{\left( n\right) }=\epsilon _{b}^{\left( n\right) }+\epsilon
_{bb}^{\left( n\right) },
\end{equation}
with $\epsilon _{b}^{\left( n\right) }=\partial f_{b}^{\left( n\right)
}/\partial \eta $ and $\epsilon _{bb}^{\left( n\right) }=\partial
f_{bb}^{\left( n\right) }/\partial \eta $. By using $f_{b}^{\left( n\right)
} $ and $f_{bb}^{\left( n\right) }$ of Eqs. (14) and (23), we can perform
the direct differentiation to obtain the corresponding internal energy
density as 
\begin{equation}
\epsilon _{b}^{\left( n\right) }=-\left( \frac{\tanh \left( \eta ^{\left(
n\right) }\right) }{4^{n}}\right) \left( \frac{\partial \eta ^{\left(
n\right) }}{\partial \eta }\right) -\sum_{k=1}^{n}\left[ \left( \frac{1}{%
4^{k}}\right) \left( \frac{\partial \eta ^{\left( k\right) }}{\partial \eta }%
\right) \right] ,
\end{equation}
and 
\begin{equation}
\epsilon _{bb}^{\left( n\right) }=-\frac{1}{2N_{b}^{\left( n\right) }}\left[
\left( \frac{\partial }{\partial \eta }\ln A_{1}^{(n)}\right) -\left( \frac{2%
}{\pi }\right) \left( \frac{\partial }{\partial \eta }\ln \kappa ^{\left(
n\right) }\right) K\left( \kappa ^{\left( n\right) }\right) \right] ,
\end{equation}
where the explicit forms of the functions, $\partial \ln
A_{1}^{(n)}/\partial \eta $ and $\partial \ln \kappa ^{\left( n\right)
}/\partial \eta $, can be obtained by direct differentiations of Eqs. (20)
and (22), 
\begin{equation}
\frac{\partial }{\partial \eta }\ln A_{1}^{(n)}=\left( \frac{\cosh \left(
\eta ^{\left( n\right) }\right) }{\sinh \left( \eta ^{\left( n\right)
}\right) }-\frac{3\sinh \left( \eta ^{\left( n\right) }\right) }{\cosh
\left( \eta ^{\left( n\right) }\right) }\right) \left( \frac{\partial \eta
^{\left( n\right) }}{\partial \eta }\right) ,
\end{equation}
and 
\begin{equation}
\frac{\partial }{\partial \eta }\ln \kappa ^{\left( n\right) }=\left( \frac{%
2\cosh \left( 2\eta ^{\left( n\right) }\right) }{\sinh \left( 2\eta ^{\left(
n\right) }\right) }-\frac{4\sinh \left( 2\eta ^{\left( n\right) }\right) }{%
\cosh \left( 2\eta ^{\left( n\right) }\right) }\right) \left( \frac{\partial
\eta ^{\left( n\right) }}{\partial \eta }\right) ,
\end{equation}
with 
\begin{equation}
\frac{\partial \eta ^{\left( k\right) }}{\partial \eta }=2^{k}\left\{
\prod\limits_{j=1}^{k}\left[ \tanh \left( 2\eta ^{\left( j-1\right) }\right) %
\right] \right\}
\end{equation}
for $1\leq k\leq n$ obtained from the map of Eq. (10), and $K\left( \kappa
^{\left( n\right) }\right) $ is the complete elliptic integral of the first
kind defined as 
\begin{equation}
K\left( \kappa ^{\left( n\right) }\right) =\int_{0}^{\pi /2}\frac{d\omega }{%
\left( 1-\kappa ^{\left( n\right) 2}\sin ^{2}\omega \right) ^{1/2}},
\end{equation}

We notice that $K\left( \kappa ^{\left( n\right) }\right) $ is singular at
the critical point $\kappa ^{\left( n\right) }=1$. As $\kappa ^{\left(
n\right) }\rightarrow 1$, it can be shown$\left[ 22\right] $ that 
\begin{equation}
K\left( \kappa ^{\left( n\right) }\right) \underset{T\rightarrow T_{c}}{=}%
-\ln \left| \frac{T-T_{c}}{T_{c}}\right| .
\end{equation}
This logarithmic divergence appears in the specific heat, and it becomes the
signature of the Ising phase transition in two-dimensions. But, by
substituting the critical condition, $\sinh 2\eta ^{\left( n\right) }\overset%
{c}{=}1$, into Eq. (37), we obtain 
\begin{equation}
\frac{\partial }{\partial \eta }\ln \kappa ^{\left( n\right) }\overset{c}{=}%
0.
\end{equation}
As a consequence of Eq. (41), we see that $\epsilon _{bb}^{\left( n\right) }$
of Eq. (35) is a continuous function of temperature $T$ even at the critical
point, $T=T_{c}$.

To have a qualitative understanding about the varitions of $\epsilon
_{b}^{\left( n\right) }$ and $\epsilon _{bb}^{\left( n\right) }$ with
respect to the decoration-level $n$, we compute the critical values of $%
\epsilon _{b}^{\left( n\right) }$ and $\epsilon _{bb}^{\left( n\right) }$
for different $n$ values. By using the critical condition of Eq. (29) for an 
$n$-lattice, we obtain the critical values of Eqs. (36) and (38) as 
\begin{equation}
\frac{\partial }{\partial \eta }\ln A_{1}^{(n)}\overset{c}{=}\left( 4-2\sqrt{%
2}\right) \left( \frac{\partial \eta ^{\left( n\right) }}{\partial \eta }%
\right) _{c},
\end{equation}
and 
\begin{equation}
\frac{\partial \eta ^{\left( k\right) }}{\partial \eta }\overset{c}{=}%
2^{k}\left\{ \prod\limits_{j=n-k+1}^{n}\left[ \frac{\left( h^{\left(
j\right) }\right) ^{4}-1}{\left( h^{\left( j\right) }\right) ^{4}+1}\right]
\right\}
\end{equation}
for $1\leq k\leq n$. Here and hereafter, we use the subscript $c$ to denote
that the quantity in the parenthesis is evaluated at the critical point of
an $n$-lattice. Then, using Eqs. (41) and (42) we can express the critical
value of $\epsilon _{b}^{\left( n\right) }$ and $\epsilon _{bb}^{\left(
n\right) }$ given by Eqs. (34) and (35) as 
\begin{equation}
\epsilon _{b}^{\left( n\right) }\overset{c}{=}-\frac{1}{4^{n}}\left( \frac{%
\sqrt{2}}{2+\sqrt{2}}\right) \left( \frac{\partial \eta ^{\left( n\right) }}{%
\partial \eta }\right) _{c}-\sum_{k=1}^{n}\left[ \left( \frac{1}{4^{k}}%
\right) \left( \frac{\partial \eta ^{\left( k\right) }}{\partial \eta }%
\right) _{c}\right] ,
\end{equation}
and 
\begin{equation}
\epsilon _{bb}^{\left( n\right) }\overset{c}{=}-\frac{1}{4^{n+1}}\left( 4-2%
\sqrt{2}\right) \left( \frac{\partial \eta ^{\left( n\right) }}{\partial
\eta }\right) _{c},
\end{equation}
with $\left( \partial \eta ^{\left( k\right) }/\partial \eta \right) _{c}$
given by Eq. (43).

The numerical results of Eqs. (44) and (45) for some decoration levels $n$
are given in Table 1, and the corresponding curves are shown in Fig. 4.
These results indicate that the local correlations among the $b^{\left(
n\right) }$ bonds of a given $n$-bond are enhenced and the long-range
correlations among different $n$-bonds are reduced as $n$ increases. For $%
n\geq 6$, the long-range correlations are almost absent, and the system is
very closed to that composing of independent $n$-bonds. \ 

In calculating the specific heat $C_{V}^{\left( n\right) }$ of an $n$%
-lattice, we use the dimensionless quantity, $c^{\left( n\right)
}=C_{V}^{\left( n\right) }/k_{B}$, defined as 
\begin{equation}
c^{(n)}=-\ \eta ^{2}\frac{\partial \epsilon ^{\left( n\right) }}{\partial
\eta }.
\end{equation}
Similar to the internal energy density, we can express $c^{(n)}$ as the sum
of two parts, 
\begin{equation}
c^{\left( n\right) }=c_{b}^{\left( n\right) }+c_{bb}^{\left( n\right) },
\end{equation}
with $c_{b}^{\left( n\right) }=-\eta ^{2}\partial \epsilon _{b}^{\left(
n\right) }/\partial \eta $ and $c_{b}^{\left( n\right) }=-\eta ^{2}\partial
\epsilon _{b}^{\left( n\right) }/\partial \eta $.

For the direct differentiation of $\epsilon _{b}^{\left( n\right) }$ of Eq.
(34) with respect to $\eta $, we obtain 
\begin{equation}
\frac{c_{b}^{\left( n\right) }}{\eta ^{2}}=\frac{1}{4^{n}}\left[ \left( 
\frac{\partial \eta ^{\left( n\right) }/\partial \eta }{\cosh \left( \eta
^{\left( n\right) }\right) }\right) ^{2}+\left( \tanh \left( \eta ^{\left(
n\right) }\right) \right) \left( \frac{\partial ^{2}\eta ^{\left( n\right) }%
}{\partial \eta ^{2}}\right) \right] +\left\{ \sum_{k=1}^{n}\left[ \left( 
\frac{1}{4^{k}}\right) \left( \frac{\partial ^{2}\eta ^{\left( k\right) }}{%
\partial \eta ^{2}}\right) \right] \right\} ,
\end{equation}
where the function, $\partial \eta ^{\left( k\right) }/\partial \eta $, is
given by Eq. (38), and its further derivative yields 
\begin{equation}
\frac{\partial ^{2}\eta ^{\left( k\right) }}{\partial \eta ^{2}}=\left( 
\frac{\partial \eta ^{\left( k\right) }}{\partial \eta }\right) \left\{
\sum\limits_{i=1}^{k}\left[ \left( \frac{2^{i+1}}{\sinh \left( 4\eta
^{\left( i-1\right) }\right) }\right) \left( \prod\limits_{j=1}^{i-1}\tanh
\left( 2\eta ^{\left( j-1\right) }\right) \right) \right] \right\}
\end{equation}
for $k\geq 1$. Note that in Eq. (49) the product for the index $j$ in the
last parenthesis exists only for $i\geq 2$.

On the other hand, the direct differentiation of $\epsilon _{bb}^{\left(
n\right) }$ of Eq. (35)\ with respect to $\eta $ yields 
\begin{equation}
\frac{c_{bb}^{\left( n\right) }}{\eta ^{2}}=\frac{1}{2N_{b}^{\left( n\right)
}}\left\{ \left( \frac{\partial ^{2}}{\partial \eta ^{2}}\ln A_{1}^{\left(
n\right) }\right) -\frac{2}{\pi }\left[ P_{0}(\kappa ^{\left( n\right)
})K\left( \kappa ^{\left( n\right) }\right) +Q_{0}\left( \kappa ^{\left(
n\right) }\right) E\left( \kappa ^{\left( n\right) }\right) \right] \right\}
,
\end{equation}
where the functions, $P_{0}\left( \kappa ^{\left( n\right) }\right) $ and $%
Q_{0}\left( \kappa ^{\left( n\right) }\right) $, are defined as 
\begin{equation}
P_{0}\left( \kappa ^{\left( n\right) }\right) =\left( \frac{\partial ^{2}}{%
\partial \eta ^{2}}\ln \kappa ^{\left( n\right) }\right) -\left( \frac{%
\partial }{\partial \eta }\ln \kappa ^{\left( n\right) }\right) ^{2}
\end{equation}
and 
\begin{equation}
Q_{0}\left( \kappa ^{\left( n\right) }\right) =\left( \frac{1}{1-\left(
\kappa ^{\left( n\right) }\right) ^{2}}\right) \left( \frac{\partial }{%
\partial \eta }\ln \kappa ^{\left( n\right) }\right) ^{2},
\end{equation}
and $E\left( \kappa ^{\left( n\right) }\right) $ is referred as the complete
elliptic integral of the second kind and defined as 
\begin{equation}
E\left( \kappa ^{\left( n\right) }\right) =\int_{0}^{\pi /2}d\omega \left(
1-\kappa ^{\left( n\right) 2}\sin ^{2}\omega \right) ^{1/2}.
\end{equation}
Note that in obtaining Eq. (50) we have used the well-known formula$\left[ 22%
\right] $, 
\begin{equation}
\frac{\partial K\left( \kappa ^{\left( n\right) }\right) }{\partial \kappa
^{\left( n\right) }}=\left( \frac{1}{\kappa ^{\left( n\right) }}\right) %
\left[ \left( \frac{1}{1-\left( \kappa ^{\left( n\right) }\right) ^{2}}%
\right) E\left( \kappa ^{\left( n\right) }\right) -K\left( \kappa ^{\left(
n\right) }\right) \right] .
\end{equation}
We also notice that by further differentiating the results of Eqs. (36) and
(37) with respect to $\eta $ we obtain 
\begin{equation}
\frac{\partial ^{2}}{\partial \eta ^{2}}\ln A_{1}^{\left( n\right)
}=-P_{1}\left( \eta ^{\left( n\right) }\right) \left( \frac{\partial \eta
^{\left( n\right) }}{\partial \eta }\right) ^{2}+Q_{1}\left( \eta ^{\left(
n\right) }\right) \left( \frac{\partial ^{2}\eta ^{\left( n\right) }}{%
\partial \eta ^{2}}\right) ,
\end{equation}
and 
\begin{equation}
\frac{\partial ^{2}}{\partial \eta ^{2}}\ln \kappa ^{\left( n\right)
}=-P_{2}\left( \eta ^{\left( n\right) }\right) \left( \frac{\partial \eta
^{\left( n\right) }}{\partial \eta }\right) ^{2}+Q_{2}\left( \eta ^{\left(
n\right) }\right) \left( \frac{\partial ^{2}\eta ^{\left( n\right) }}{%
\partial \eta ^{2}}\right) ,
\end{equation}
with 
\begin{equation}
P_{1}\left( \eta ^{\left( n\right) }\right) =\frac{1}{\sinh ^{2}\left( \eta
^{\left( n\right) }\right) }+\frac{3}{\cosh ^{2}\left( \eta ^{\left(
n\right) }\right) },
\end{equation}
\begin{equation}
Q_{1}\left( \eta ^{\left( n\right) }\right) =\frac{\cosh \left( \eta
^{\left( n\right) }\right) }{\sinh \left( \eta ^{\left( n\right) }\right) }-%
\frac{3\sinh \left( \eta ^{\left( n\right) }\right) }{\cosh \left( \eta
^{\left( n\right) }\right) },
\end{equation}
\begin{equation}
P_{2}\left( \eta ^{\left( n\right) }\right) =\frac{4}{\sinh ^{2}\left( 2\eta
^{\left( n\right) }\right) }+\frac{8}{\cosh ^{2}\left( 2\eta ^{\left(
n\right) }\right) },
\end{equation}
and 
\begin{equation}
Q_{2}\left( \eta ^{\left( n\right) }\right) =\frac{2\cosh \left( 2\eta
^{\left( n\right) }\right) }{\sinh \left( 2\eta ^{\left( n\right) }\right) }-%
\frac{4\sinh \left( 2\eta ^{\left( n\right) }\right) }{\cosh \left( 2\eta
^{\left( n\right) }\right) }.
\end{equation}

The valuess of $c_{b}^{\left( n\right) }$ and $c_{bb}^{\left( n\right) }$ of
Eqs. (48) and (50) can be calculated by using the explicit forms of Eqs.
(36-38), (49), (55), and (56). The numerical results as a function of
reduced temperature $k_{B}T/J$ for some $n$ values are shown in Fig. 5. The
peak temperature of $c_{b}^{\left( n\right) }$, denoted by $\left(
k_{B}T_{m}/J\right) _{n}$, marks the occurrence of the local ordering in an $%
n$-bond, and the numerical values of $\left( k_{B}T_{m}/J\right) _{n}$ for
some decoration levels $n$ are given in Table 1. The plot of $\left(
k_{B}T_{m}/J\right) _{n}$ versus $n$ is also shown in the lower part of Fig.
3 with the solid curve given by the fitting function $\left(
k_{B}T_{m}/J\right) _{n}=1.6410-\left( 0.8063\right) \exp \left[ -\left(
0.7144\right) n\right] $. Thus, because of the enhancement of the local
correlation for the increase of $n$, $\left( k_{B}T_{m}/J\right) _{n}$
increases as $n$ increases. Moreover, the peak temperature of $%
c_{bb}^{\left( n\right) }$ is the critical temperature $\left(
k_{B}T_{c}/J\right) _{n}$ which signifies the occurrence of the long-range
ordering among the Ising spins defined on the primary sites, and the
behavior of $\left( k_{B}T_{c}/J\right) _{n}$ versus $n$ is already given in
the upper part of Fig. 3. The results of Fig. 3 indicate that the local
ordering occurs at the temperature below the critical temperature, and this
is exactly opposite to the case of cell-decorated triangular Ising model$%
\left[ 14\right] $.

The singular behavior of $c^{\left( n\right) }$ arises from $c_{bb}^{\left(
n\right) }$ of Eq. (50) in the form of the logarithmic divergence given by
Eq. (40). Thus, we can express the singular behavior of $c^{\left( n\right)
} $ as 
\begin{equation}
c^{\left( n\right) }\underset{T\rightarrow T_{c}}{=}-A_{\sin g}^{\left(
n\right) }\ln \left| \frac{T-T_{c}}{T_{c}}\right| .
\end{equation}
where $A_{\sin g}^{\left( n\right) }$, referred as the critical amplitude,
is given as 
\begin{equation}
A_{\sin g}^{\left( n\right) }=-\frac{\eta _{c}^{2}}{N_{b}^{\left( n\right)
}\pi }\left[ P_{0}\left( \kappa ^{\left( n\right) }\right) \right] _{c},
\end{equation}
with $P_{0}\left( \kappa ^{\left( n\right) }\right) $ given by Eq. (51). By
using Eq. (41), we can rewrite the expression of $A_{\sin g}^{\left(
n\right) }$ as 
\begin{equation}
A_{\sin g}^{\left( n\right) }=-\frac{\eta _{c}^{2}}{N_{b}^{\left( n\right)
}\pi }\left( \frac{\partial ^{2}}{\partial \eta ^{2}}\ln \kappa ^{\left(
n\right) }\right) _{c}.
\end{equation}
By taking the critical value of Eq. (56) with $\left[ Q_{2}\left( \eta
^{\left( n\right) }\right) \right] _{c}=0$ and $\left[ P_{2}\left( \eta
^{\left( n\right) }\right) \right] _{c}=8$, we can use the result of Eq.
(43) to obtain 
\begin{equation}
\frac{\partial ^{2}}{\partial \eta ^{2}}\ln \kappa ^{\left( n\right) }%
\overset{c}{=}-\frac{4^{n+2}}{2}\left\{ \prod\limits_{k=1}^{n}\left[ \frac{%
\left( h^{\left( k\right) }\right) ^{4}-1}{\left( h^{\left( k\right)
}\right) ^{4}+1}\right] \right\} ^{2}.
\end{equation}
Thus, using the reduced critical temperature given by Eq. (31) we can
express Eq. (63) as 
\begin{equation}
A_{\sin g}^{\left( n\right) }=\frac{4}{\pi }\left[ \ln h^{\left( n\right) }%
\right] ^{2}\left\{ \prod\limits_{k=1}^{n}\left[ \frac{\left( h^{\left(
k\right) }\right) ^{4}-1}{\left( h^{\left( k\right) }\right) ^{4}+1}\right]
\right\} ^{2}.
\end{equation}

The numerical values of Eq. (65) for some $n$ values are given in Table 1.
The plot of $A_{\sin g}^{\left( n\right) }$ versus $n$ is also given in Fig.
6 where the solid curve is given by the fitting function $A_{\sin g}^{\left(
n\right) }=\left( 0.2473\right) \exp \left[ -\left( 0.3018\right) n\right] $%
. Since the critical amplitude associated with the logrithmic singularity
characterizes the width of the critical region, our results indicate that as
increasing $n$ the critical region is narrowing down to the vanish limit.
This characteristic feature in the specific heat can also be seen in the
model on anisotropic regular lattices$\left[ 13,14,20,22\right] $. We also
notice that for the cell-decorated triangular Ising model the critical
amplitude behaves as $A_{\sin g}^{\left( n\right) }\thicksim 3^{-n}\left(
n\ln n\right) ^{2}$ for very large $n\left[ 14\right] $. Thus, as the
decoration level $n$ increases the critical amplitude approaches zero less
quickly for the bond-decorated model in comparing with the cell-decorated
model.

\section{infinite decoration-level}

In this section, we discuss the critical properties of the Ising model on an 
$n$-lattice in the limit of infinite $n$. First we demonstrate the absence
of Ising phase transition. This feature is caused mainly by the
fragmentation of $n$-bonds. Then, we calculate the critical values of the
internal energy and specific heat. The specific heat curve near the critical
point becomes a cusp, and the exact critical value is given.

The map of Eq. (10) has three fixed points, $\eta _{f}$, given by the
solutions of 
\begin{equation}
\eta _{f}=\ln \left[ \cosh \left( 2\eta _{f}\right) \right] .
\end{equation}
Among the three fixed points, two are attractors locating at $\eta
_{f}^{a_{1}}=0$ and $\eta _{f}^{a_{2}}=\infty $ respectively, and one is a
repellor locating at $\eta _{f}^{r}=0.609378$. For the limit of infinite
decoration-level, the repellor is the critical point of the bulk system.
Since the critical point is a repellor, the critical behavior of the system
may be different from that obtained by extrapolating the results of
finite-decorated systems to the limit of infinite $n$. To show this, first
we give the plots of $\eta ^{\left( n\right) }$ versus $\eta $ for $n=0$, $2$%
, $4$, and $6$ in Fig. 7. The results of Fig. 7 indicate that for the limit
of infinite $n$ the possible $\eta ^{\left( n\right) }$ values are
restricted to the fixed points of Eq. (66): $\eta ^{\left( n\right) }=\eta
_{f}^{a_{1}}$ for $\eta <\eta _{f}^{r}$, $\eta ^{\left( n\right) }=\eta
_{f}^{r}$ for $\eta =\eta _{f}^{r}$, and $\eta ^{\left( n\right) }=\eta
_{f}^{a_{2}}$ for $\eta >\eta _{f}^{r}$. This leads to the result that the
value of $\kappa ^{\left( n\right) }$ defined by Eq. (22) has only two
possible values, 
\begin{equation}
\kappa ^{\left( \infty \right) }=0
\end{equation}
for $\eta \lessgtr \eta _{f}^{r}$, and 
\begin{equation}
\kappa ^{\left( \infty \right) }=2\sqrt{\left[ \exp (-2\eta _{f}^{r})\right]
-\left[ \exp \left( -4\eta _{f}^{r}\right) \right] }
\end{equation}
for $\eta =\eta _{f}^{r}$. The latter gives the numerical value $\kappa
^{\left( \infty \right) }=0.912622$. Thence, the logarithmic divergence of
the Ising transition, caused by the divergence of the elliptic integral $%
K\left( \kappa ^{\left( n\right) }\right) $ of Eq. (39) at the point $\kappa
^{\left( n\right) }=1$, is absent for the infinite $n$ limit. This indicates
that the cross over from a finite-decorated system to an infinite-decorated
system may not be a smooth continuation.

To exhibit the critical properties of the infinite-decorated system more
explicitly, we then calculate the critical values of the internal energy and
specific heat. Since the critical point of the infinite-decorated system is
the repellor of the map of Eq. (10), we have the following identities for
the derivatives evaluated at the critical point $\eta =\eta _{f}^{r}$: 
\begin{equation}
\frac{\partial \eta ^{\left( i+k\right) }}{\partial \eta ^{\left( i\right) }}%
\overset{\ast }{=}\frac{\partial \eta ^{\left( k\right) }}{\partial \eta },
\end{equation}
and 
\begin{equation}
\frac{\partial ^{2}\eta ^{\left( i+k\right) }}{\partial \eta ^{\left(
i\right) 2}}\overset{\ast }{=}\frac{\partial ^{2}\eta ^{\left( k\right) }}{%
\partial \eta ^{2}}.
\end{equation}
Here, we use the notation, $\overset{\ast }{=}$ , to denote the equivalence
established at the repellor. On the other hand, by successively applying the
chain rule we have 
\begin{equation}
\frac{\partial \eta ^{\left( k\right) }}{\partial \eta }=\prod_{i=1}^{k}%
\frac{\partial \eta ^{\left( i\right) }}{\partial \eta ^{\left( i-1\right) }}%
.
\end{equation}
Then, as a result of Eq. (69) we obtain 
\begin{equation}
\frac{\partial \eta ^{\left( k\right) }}{\partial \eta }\overset{\ast }{=}%
\left( d^{\left( 1\right) }\right) ^{k},
\end{equation}
with $d^{\left( 1\right) }=\left( \partial \eta ^{\left( 1\right) }/\partial
\eta \right) _{\eta =\eta _{f}^{r}}=1.678574$.

For the second derivative, $\partial ^{2}\eta ^{\left( k\right) }/\partial
\eta ^{2}$, we first perform the direct differentiation of Eq. (71) and then
use the chain rule to obtain 
\begin{equation}
\frac{\partial ^{2}\eta ^{\left( k\right) }}{\partial \eta ^{2}}=\left( 
\frac{\partial ^{2}\eta ^{\left( 1\right) }}{\partial \eta ^{2}}\right) %
\left[ D_{2}^{k}\left( \eta \right) \right] +\left( \frac{\partial \eta
^{\left( 1\right) }}{\partial \eta }\right) ^{2}\left( \frac{\partial
^{2}\eta ^{\left( 2\right) }}{\partial \eta ^{\left( 1\right) 2}}\right) %
\left[ D_{3}^{k}\left( \eta \right) \right] +...+\left[ D_{1}^{k-1}\left(
\eta \right) \right] ^{2}\left( \frac{\partial ^{2}\eta ^{\left( k\right) }}{%
\partial \eta ^{\left( k-1\right) 2}}\right) ,
\end{equation}
with 
\begin{equation}
D_{a}^{k}\left( \eta \right) =\prod_{i=a}^{k}\frac{\partial \eta ^{\left(
i\right) }}{\partial \eta ^{\left( i-1\right) }}
\end{equation}
for $a\leq k$. Then, by using the identities of Eqs. (69) and (70) we can
express Eq. (73) as 
\begin{equation}
\frac{\partial ^{2}\eta ^{\left( k\right) }}{\partial \eta ^{2}}\overset{%
\ast }{=}d^{\left( 2\right) }\left[ \left( d^{\left( 1\right) }\right)
^{k-1}+\left( d^{\left( 1\right) }\right) ^{k}+...+\left( d^{\left( 1\right)
}\right) ^{2\left( k-1\right) }\right]
\end{equation}
which yields 
\begin{equation}
\frac{\partial ^{2}\eta ^{\left( k\right) }}{\partial \eta ^{2}}\overset{%
\ast }{=}d^{\left( 2\right) }\left( d^{\left( 1\right) }\right) ^{k-1}\left( 
\frac{\left( d^{\left( 1\right) }\right) ^{k}-1}{d^{\left( 1\right) }-1}%
\right) ,
\end{equation}
with $d^{\left( 2\right) }=\left( \partial ^{2}\eta ^{\left( 1\right)
}/\partial \eta ^{2}\right) _{\eta =\eta _{f}^{r}}=1.182391$.

To calculate the internal energy density and specific heat at the repellor,
we take $\eta =\eta _{f}^{r}$, Eq. (72) for $\left( \partial \eta ^{\left(
n\right) }/\partial \eta \right) _{\eta =\eta _{f}^{r}}$, and Eq. (76) for $%
\left( \partial ^{2}\eta ^{\left( n\right) }/\partial \eta ^{2}\right)
_{\eta =\eta _{f}^{r}}$. Then, for the internal energy density of Eqs. (34)
and (35) we obtain the numerical results as 
\begin{equation}
\epsilon _{b}^{\left( n\right) }\overset{\ast }{=}-\left( 0.543689\right)
\left( a^{\left( 1\right) }\right) ^{n}-a^{\left( 1\right) }\left( \frac{%
1-\left( a^{\left( 1\right) }\right) ^{n}}{1-a^{\left( 1\right) }}\right) ,
\end{equation}
and 
\begin{equation}
\epsilon _{bb}^{\left( n\right) }\overset{\ast }{=}-\left( 0.414742\right)
\left( a^{\left( 1\right) }\right) ^{n},
\end{equation}
with $a^{\left( 1\right) }=d^{\left( 1\right) }/4<1$. Similarly, for the
specific heat of Eqs. (48) and (50) at the repellor we obtain the numerical
results as 
\begin{equation}
c_{b}^{\left( n\right) }\overset{\ast }{=}\left( 0.261574\right) \left(
a^{\left( 2\right) }\right) ^{n}+\left( 0.201894\right) g^{\left( 1\right)
}\left( 1-\frac{1}{\left( d^{\left( 1\right) }\right) ^{n}}\right) \left(
a^{\left( 2\right) }\right) ^{n}+\left( 0.371342\right) g^{\left( 1\right) }%
\left[ g^{\left( 2\right) }-g^{\left( 3\right) }\right] ,
\end{equation}
and 
\begin{equation}
c_{bb}^{\left( n\right) }\overset{\ast }{=}-\left( 0.115082\right) \left(
a^{\left( 2\right) }\right) ^{n}+\left( 0.154011\right) g^{\left( 1\right)
}\left( 1-\frac{1}{\left( d^{\left( 1\right) }\right) ^{n}}\right) \left(
a^{\left( 2\right) }\right) ^{n},
\end{equation}
with $a^{\left( 2\right) }=\left( d^{\left( 1\right) }\right) ^{2}/4<1$, $%
g^{\left( 1\right) }=d^{\left( 2\right) }/\left( \left( d^{\left( 1\right)
}\right) ^{2}-d^{\left( 1\right) }\right) $, $g^{\left( 2\right) }=a^{\left(
2\right) }\left( 1-\left( a^{\left( 2\right) }\right) ^{n}\right) /\left(
1-a^{\left( 2\right) }\right) $, and $g^{\left( 3\right) }=a^{\left(
1\right) }\left( 1-\left( a^{\left( 1\right) }\right) ^{n}\right) /\left(
1-a^{\left( 1\right) }\right) $.

For the limit of infinite decoration-level, we have $\left( a^{\left(
1\right) }\right) ^{n}\rightarrow 0$, $\left( a^{\left( 2\right) }\right)
^{n}\rightarrow 0$, and $1/\left( d^{\left( 1\right) }\right)
^{n}\rightarrow 0$ as $n\rightarrow \infty $. Then, in this limit \ Eqs.
(77) and (79) reduce to 
\begin{equation}
\epsilon _{b}^{\left( \infty \right) }\overset{\ast }{=}-\left( \frac{%
a^{\left( 1\right) }}{1-a^{\left( 1\right) }}\right) =-0.723079,
\end{equation}
and 
\begin{equation}
c_{b}^{\left( \infty \right) }\overset{\ast }{=}\left( 0.371342\right)
g^{\left( 1\right) }\left[ \left( \frac{a^{\left( 2\right) }}{1-a^{\left(
2\right) }}\right) -\left( \frac{a^{\left( 1\right) }}{1-a^{\left( 1\right) }%
}\right) \right] =0.639852,
\end{equation}
while the long-range parts vanish, $\epsilon _{bb}^{\left( \infty \right) }%
\overset{\ast }{=}0$ and $c_{bb}^{\left( \infty \right) }\overset{\ast }{=}0$

From these results we may conclude that the system of an $n$-lattice reduces
to that composing of independent $n$-bonds as $n\rightarrow \infty $. In
this limit, our results also show that the critical specific heat becomes a
cusp with the height $c^{\left( \infty \right) }=0.639852$ at the critical
point. Thus, the corresponding critical exponent of the specific heat, $%
\alpha $, is the same as that of the diamond-hierarchical Ising model, $%
\alpha =$ $-0.67652\left[ 9,23\right] $. \ 

\section{summary and discussion}

We use the Ising model defined on square lattices with diamond-type
bond-decorations to study the ferromagnetic phase transition of an
inhomogeneous system. In this model, the decoration level $n$ associated
with a decorated lattice can be used as a parameter to characterize the
degree of the inhomogeneity of a system. Based on the exact solutions of the
two-dimensional Ising mobel, we employ the bond-renormalization scheme to
obtain the analytic form of the free energy. The free energy and its derived
quantities, the internal energy and specific heat, are expressed as the sum
of those contributed by the local interactions among the sub-bonds of an $n$%
-bond and the long-range interactions among different $n$-bonds. Then, the
change in the nature of the phase transition as $n$ increases can be viewed
as the reflection of the competetion between the local and the long-range
interactions.

In general, as $n$ increases, the long-range interaction is reduced and the
local interaction is enhenced. The logarithmic singularity of the specific
heat at the critical point is caused by the long-range interaction. Along
with the reduction of the long-range interaction as $n$ gets larger, we have
the critical temperature lowering down and the width of the critical region
becoming narrower. By fitting the data numerically, we obtain the critical
temperature changes with the decoration level $n$ as $\left(
k_{B}T_{c}/J\right) _{n}=1.6410+\left( 0.6281\right) \exp \left[ -\left(
0.5857\right) n\right] $ and the critical amplitude of the specific heat
changes as $A_{\sin g}^{\left( n\right) }=\left( 0.2465\right) 4^{n}\exp %
\left[ -\left( 1.6530\right) n\right] $. Moreover, because of the
enhancement of the local interactions, the occurring temperature of the
local ordering increases with the form, $\left( k_{B}T_{m}/J\right)
_{n}=1.6410-\left( 0.1081\right) \exp \left[ -\left( 0.4098\right) n\right] $%
, as $n$ increases. Both $\left( k_{B}T_{c}/J\right) _{n}$ and $\left(
k_{B}T_{m}/J\right) _{n}$ approach to the same limit $1.6410$ as $n$
approaches the infinite limit. Thus, the distance between $\left(
k_{B}T_{c}/J\right) _{n}$ and $\left( k_{B}T_{m}/J\right) _{n}$ decreases
with increasing $n$.

The cross over from a finite-decorated system to an infinite-decorated
system is not a smooth continuation. For the infinite decoration-level, the
critical temperature $k_{B}T_{c}/J$ is $1.64101(8)$ which is exactly the
location of the repellor of the renormalization map of the variable $\eta
^{\left( n\right) }$. In this limit, the system of an $n$-lattice reduces to
that of independent $n$-bonds, and the specific heat becomes a cusp with the
value $0.63985(0)$ at the critical point instead of the logarithmic
divergence.

It is interesting to compare the results we obtain with those from the
cell-decorated triangular Ising model given in Ref. $\left[ 14\right] $. The
critical region decreases with increasing $n$ for both cases. But, there
exists some different features as the followings: $\left( i\right) $ The
decreasing speed of the critical amplitude of the specific heat with
increasing $n$ is slower for the bond-decoration case in comparing with that
of the cell-decoration. $\left( ii\right) $ For the case of
bond-decorations, the occurring temperature of the local ordering $\left(
k_{B}T_{m}/J\right) _{n}$ of an $n$-lattice is smaller than the critical
temperature $\left( k_{B}T_{c}/J\right) _{n}$ and the distance between $%
\left( k_{B}T_{m}/J\right) _{n}$ and $\left( k_{B}T_{c}/J\right) _{n}$
decreases with increasing $n$. However, for the case of cell-decorations, $%
\left( k_{B}T_{m}/J\right) _{n}$ is larger than $\left( k_{B}T_{c}/J\right)
_{n}$ and $\left( k_{B}T_{m}/J\right) _{n}$ is freezed for $n\geq 5$. $%
\left( iii\right) $ There is a finite peak arising to the right of the
critical point for the specific heat due to the local ordering in the
cell-decoration case, but such a peak is not observed in the bond-decoration
case.

We may use the idea of the order of ramification to explain the differences
described in the above. To characterize the geometric difference between
bond-decorated and cell-decorated lattices, we introduce the quantities, $%
R_{1}$and $R_{2}$, with $R_{1}$ for the cutting-bond number of isolating an $%
n$-bond or $n$-cell and $R_{2}$ for the average of the total bond number
contained in an $n$-bond or $n$-cell. We notice that the $R_{1}$ value
signifies the ability of establishing the long-range correlations among
different $n$-bonds or $n$-cells and the interactions among $R_{2}$ bonds
are referred as the local interactions.

For the cell-decorated triangular Ising model with decoration-level $n$, we
have $R_{1}=12$ and $R_{2}=3^{n+1}$. Because of $R_{1}\ll R_{2}$ for $n\geq
3 $, the long-range correlation is very hard to establish, and this leads to
the higher temperature for the occurrence of the local ordering in comparing
with that of the long-range ordering. For sufficiently large $n$ with $%
R_{1}/R_{2}\simeq 0$, an $n$-cell almost become a closed subsystem, and
thence a round peak, contributed by the subsystem, appears above the
critical point in the curve of the specific heat.

For the diamond-type bond-decorated Ising model discussed in this work, we
have $R_{1}=6\cdot 2^{n}$ and $R_{2}=4^{n}$. The values of $R_{1}$ and $%
R_{2} $ are comparable for $n\leq 5$, and the long-range effect is almost
absent for $n\geq 6$. Thence, the round peak in the specific heat, which
appears in the case of cell-decorations, is absent. Moreover, the
coordination numbers of the decorated sites in an $n$-bond are highly
inhomogeneous, and this is mainly responsible for the lower occurring
temperature of the local ordering in comparing with that of the long-range
ordering.

Thus, the non-universal features of phase transition for the diamond-type
bond-decorated Ising model is quite different from that for the
cell-decorated triangular Ising model.

\section{acknowledgement}

This work was partially suppoted by the National Science Council of ROC (
Taiwan ) under the Grand No. NSC 90-2112-M-033-002.

Table 1. The numerical results of the crical temperature $\left(
k_{B}T_{c}/J\right) _{n}$, the occurring temperature of the local ordering $%
\left( k_{B}T_{m}/J\right) _{n}$, the critical values of the internal energy
density $\left( \epsilon ^{\left( n\right) }\right) _{c}$, the internal
energy density due to the local interactions $\left( \epsilon _{b}^{\left(
n\right) }\right) _{c}$, and the internal energy density due to the
long-range interactions $\left( \epsilon _{bb}^{\left( n\right) }\right) _{c}
$, and the critical amplitude at the logarithmic divergence in the specific
heat $A_{\sin g}^{\left( n\right) }.$

\begin{tabular}{ccccccc}
n & $(kT_{c}/J)_{n}$ & $(kT_{m}/J)_{n}$ & $\left( \varepsilon ^{(n)}\right)
_{c}$ & $\left( \varepsilon _{b}^{(n)}\right) _{c}$ & $\left( \varepsilon
_{bb}^{(n)}\right) _{c}$ & $A_{\sin g}$ \\ \hline\hline
0 & 2.269185 & 0.833520 & -0.707107 & -0.414214 & -0.292893 & 0.247269 \\ 
1 & 1.982072 & 1.255390 & -0.653281 & -0.541190 & -0.112085 & 0.189850 \\ 
2 & 1.833652 & 1.449949 & -0.658949 & -0.614280 & -0.044674 & 0.140957 \\ 
3 & 1.752268 & 1.540593 & -0.675920 & -0.657720 & -0.018202 & 0.102495 \\ 
4 & 1.706095 & 1.586440 & -0.691315 & -0.683810 & -0.007508 & 0.073587 \\ 
5 & 1.679371 & 1.610705 & -0.702627 & -0.699510 & -0.003119 & 0.052430 \\ 
6 & 1.663721 & 1.623980 & -0.710252 & -0.708950 & -0.001301 & 0.037184 \\ 
7 & 1.654492 & 1.631345 & -0.715166 & -0.714626 & -0.000544 & 0.026299 \\ 
8 & 1.649027 & 1.635501 & -0.718250 & -0.718022 & -0.000228 & 0.018570 \\ 
9 & 1.645783 & 1.637855 & -0.720154 & -0.720055 & -9.54965(-5) & 0.013100 \\ 
10 & 1.643854 & 1.639191 & -0.721316 & -0.721280 & -4.00447(-5) & 0.009235
\\ 
11 & 1.642707 & 1.639962 & -0.722020 & -0.722003 & -1.67971(-5) & 0.006509
\\ 
12 & 1.642024 & 1.640400 & -0.722444 & -0.722433 & -7.04693(-6) & 0.004586
\\ 
13 & 1.641617 & 1.640664 & -0.722699 & -0.722697 & -2.95673(-6) & 0.003231
\\ 
14 & 1.641375 & 1.640809 & -0.722852 & -0.722849 & -1.24066(-6) & 0.002276
\\ 
15 & 1.641231 & 1.640893 & -0.722943 & -0.722939 & -5.20605(-7) & 0.001604
\\ 
16 & 1.641145 & 1.640946 & -0.722998 & -0.723000 & -2.18461(-7) & 0.001130
\\ 
17 & 1.641093 & 1.640975 & -0.723030 & -0.723030 & -9.1674(-8) & 7.95678(-4)
\\ 
18 & 1.641063 & 1.640993 & -0.723050 & -0.723050 & -3.84699(-8) & 5.60485(-4)
\\ 
19 & 1.641045 & 1.641003 & -0.723061 & -0.723060 & -1.61435(-8) & 3.9481(-4)
\\ 
20 & 1.641034 & 1.641009 & -0.723068 & -0.723070 & -6.7745(-9) & 2.78106(-4)%
\end{tabular}

\newpage

\FRAME{ftbpFU}{7.0768in}{2.7198in}{0pt}{\Qcb{ Examples of $n$-bonds
specified by the two lattice sites $\protect\mu $ and $\protect\nu $ for (a) 
$n=0$, (b) $n=1$, and (c) $n=2$. Note that the Ising spins except $\protect%
\sigma _{\protect\mu }$ and $\protect\sigma _{\protect\nu }$ are referred as
the inner-spins.}}{}{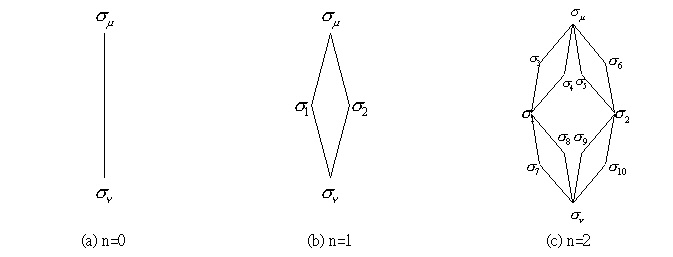}{\special{language "Scientific Word";type
"GRAPHIC";maintain-aspect-ratio TRUE;display "USEDEF";valid_file "F";width
7.0768in;height 2.7198in;depth 0pt;original-width 7.0102in;original-height
2.6775in;cropleft "0";croptop "1";cropright "1";cropbottom "0";filename
'1.jpg';file-properties "XNPEU";}}\FRAME{ftbpFU}{6.8035in}{1.6613in}{0pt}{%
\Qcb{ Examples of $n$-lattices for (a) $n=0$, (b) $n=1$, and (c) $n=2$.}}{}{%
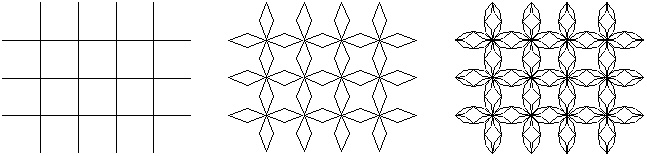}{\special{language "Scientific Word";type
"GRAPHIC";maintain-aspect-ratio TRUE;display "USEDEF";valid_file "F";width
6.8035in;height 1.6613in;depth 0pt;original-width 6.7395in;original-height
1.625in;cropleft "0";croptop "1";cropright "1";cropbottom "0";filename
'2.jpg';file-properties "XNPEU";}}

\FRAME{ftbpFU}{4.0283in}{3.4316in}{0pt}{\Qcb{The critical temperature $%
\left( k_{B}T_{c}/J\right) _{n}$ and the peak temperature of the specific
heat $\left( k_{B}T_{m}/J\right) _{n}$ caused by the local interactios of an 
$n$-lattice denoted by black dots and black triangles respectively. The
continuous curves are given by the fitting functions as $\left(
k_{B}T_{c}/J\right) _{n}=1.6410+\left( 0.6281\right) \exp \left[ -\left(
0.5857\right) n\right] $ and $\left( k_{B}T_{m}/J\right) _{n}=1.6410-\left(
0.8063\right) \exp \left[ -\left( 0.7144\right) n\right] $ respectively.}}{}{%
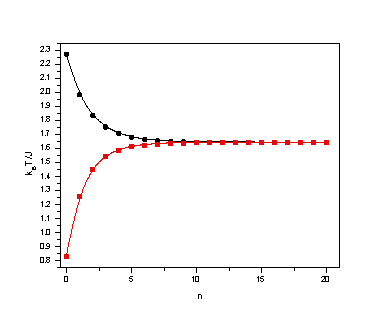}{\special{language "Scientific Word";type
"GRAPHIC";maintain-aspect-ratio TRUE;display "USEDEF";valid_file "F";width
4.0283in;height 3.4316in;depth 0pt;original-width 3.979in;original-height
3.3857in;cropleft "0";croptop "1";cropright "1";cropbottom "0";filename
'3.jpg';file-properties "XNPEU";}}\FRAME{ftbpFU}{4.1329in}{3.4108in}{0pt}{%
\Qcb{The critical values of $\protect\epsilon ^{\left( n\right) }$
(triangles), $\protect\epsilon _{b}^{\left( n\right) }$ (black dots), and $%
\protect\epsilon _{bb}^{\left( n\right) }$ (squares) as a function of
decoration-level $n$.}}{}{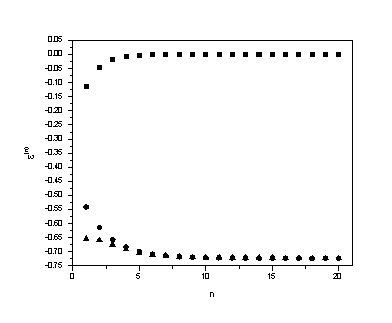}{\special{language "Scientific Word";type
"GRAPHIC";maintain-aspect-ratio TRUE;display "USEDEF";valid_file "F";width
4.1329in;height 3.4108in;depth 0pt;original-width 4.0836in;original-height
3.365in;cropleft "0";croptop "1";cropright "1";cropbottom "0";filename
'4.jpg';file-properties "XNPEU";}}\FRAME{ftbpFU}{4.0914in}{3.4523in}{0pt}{%
\Qcb{The numerical results of $c_{b}^{\left( n\right) }$ (with round peaks)
and $c_{bb}^{\left( n\right) }$ (with sharp peaks) as a function of reduced
temperature $k_{B}T/J$ for $n=2$ , $4$, and $6$.}}{}{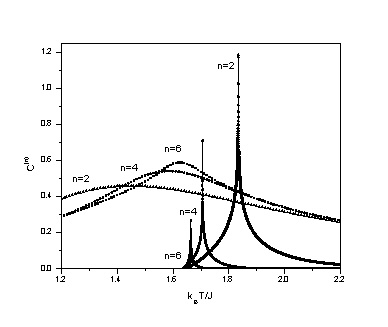}{\special%
{language "Scientific Word";type "GRAPHIC";maintain-aspect-ratio
TRUE;display "USEDEF";valid_file "F";width 4.0914in;height 3.4523in;depth
0pt;original-width 4.0413in;original-height 3.4065in;cropleft "0";croptop
"1";cropright "1";cropbottom "0";filename '5.jpg';file-properties "XNPEU";}}

\FRAME{ftbpFU}{4.0387in}{3.3797in}{0pt}{\Qcb{The critical amplitude $A_{\sin
g}^{\left( n\right) }$ of the specific heat as a function of
decoration-level $n$ with the solid curve given by the fitting function $%
A_{\sin g}^{\left( n\right) }=\left( 0.2473\right) \exp \left[ -\left(
0.3018\right) n\right] $.}}{}{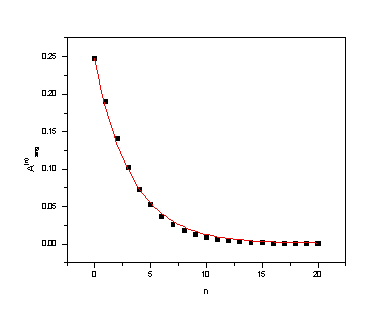}{\special{language "Scientific
Word";type "GRAPHIC";maintain-aspect-ratio TRUE;display "USEDEF";valid_file
"F";width 4.0387in;height 3.3797in;depth 0pt;original-width
3.9894in;original-height 3.333in;cropleft "0";croptop "1";cropright
"1";cropbottom "0";filename '6.jpg';file-properties "XNPEU";}}

\FRAME{ftbpFU}{4.0811in}{3.4627in}{0pt}{\Qcb{The variable $\protect\eta %
^{(n)}$ as a function of $\protect\eta $ for $n=0$, $2$, $4$, and $6$. Note
that the vertical line corresponds to the location of the repellor of the
map $\protect\eta ^{\left( n-1\right) }\rightarrow \protect\eta ^{\left(
n\right) }$.}}{}{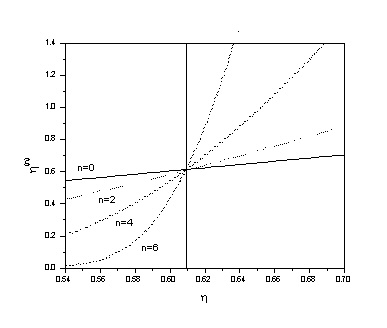}{\special{language "Scientific Word";type
"GRAPHIC";maintain-aspect-ratio TRUE;display "USEDEF";valid_file "F";width
4.0811in;height 3.4627in;depth 0pt;original-width 4.0309in;original-height
3.4169in;cropleft "0";croptop "1";cropright "1";cropbottom "0";filename
'7.jpg';file-properties "XNPEU";}}

\newpage 

\end{document}